# Investigation of Bioglass-Electrode Interfaces after Thermal Poling

C.R. Mariappan, B. Roling[*]

*Fachbereich Chemie, Physikalische Chemie, Philipps-Universität Marburg,
Hans-Meerwein-Straße, 35032 Marburg, Germany*

**Abstract**

Electrical and electrochemical processes in a bioactive soda-lime phosphosilicate glasses and in a bioabsorbable soda-lime phosphate glass during thermal poling were studied by means of thermally stimulated depolarization current measurements, ac impedance spectroscopy, and SEM/EDX analyses. The thermal poling was done by sputtering thin Pt electrode films onto the faces of the glass samples and by applying voltages up to 1 kV to the electrodes at temperatures up to 513 K. The poling leads to the formation of interfacial layers under the electrodes which are responsible for two depolarization current peaks and for one additional semicircle in a Nyquist plot of the ac impedance. The SEM/EDX profiles suggest that redox and transport processes of $Na^+$ ions are responsible for the formation of the interfacial layers and that $Ca^{2+}$ ions are immobile under the poling conditions. The sodium depletion layer under the anode leads to a negative surface charge of the glass samples which may enhance their bioactivity.

[*] – Corresponding author: B. Roling, Phone: +49 6421 28 22310, Fax: +49 6421 28 22309, e-mail: roling@staff.uni-marburg.de

## 1. Introduction

Several poling techniques have been used to enhance materials properties, such as optical non-linearity, breaking strength, chemical durability, and bioactivity [1-15]. Among these, thermal poling is one of the so-called thermally stimulated polarization techniques [6,7]. In this technique, a dc voltage $V_p$ is applied to the material for a time interval $t_p$ at a temperature $T_p$. The experimental conditions are chosen such that the applied voltage causes reorientation of dipolar units and/or movements of mobile charge carriers. Then the material is cooled to a lower temperature under the applied dc voltage. Thereby, the dipolar units and charge carriers are 'frozen' in their positions. The resulting static polarization may have a strong influence on the physical and chemical properties of the material.

Recently, it was discovered that the thermal poling technique can be used to accelerate and decelerate bone-like apatite formation on the surfaces of hydroxyapatite $Ca_5(PO_4)_3OH$ and of bioglass $46.1SiO_2 - 24.4Na_2O - 26.9CaO - 2.6P_2O_5$ (46S5) [10-12]. The influence of surface charges on the bioactivity of hydroxyapatite (HAP) was confirmed by the overgrowth of calcium phosphate layers in simulated body fluids (SBF), by the proliferation of certain cells in physiological fluids, and by an enhanced osteoconductivity in the body. When poled HAP samples were placed in SFB (which is equal to human blood plasma with the following ionic concentrations: $Na^+$ 142.0 mM, $K^+$ 5.0 mM, $Ca^{2+}$ 2.5 mM, $Mg^{2+}$ 1.5 mM, $Cl^-$ 147.8 mM, $HCO_3^-$ 4.2 mM, $HPO_4^{2-}$ 1.0 mM, $SO_4^{2-}$ 0.5mM, and with a pH of 7.25-7.40), calcium ions from the SFB were adsorbed readily on the negatively charged surface and followed by $PO_4^{3-}$, resulting in an accelerated formation of bone-like apatite [5]. The 46S5 bioglass achieved much higher polarizations with 100 times lower dc fields than the HAP. After polarizing the 46S5 glass at temperatures of about 723 K for about 1 h and after subsequent cooling to room temperature, the surface charges persist long enough to result in a significant enhancement of the bioactivity.

Up to now, the electrical and electrochemical processes leading to the formation of surface charges in bioactive materials and the chemical nature of the electrode-biomaterial interfaces have not been studied in detail. In our view, this is not only important from a basic science point of view, but is also relevant for applications of bioglasses. If, for instance, the negative surface charge of a biomaterial after poling was caused by a frozen-in electrochemical double layer extending only over a few atomic distances, the negative charge would be easily removable by mechanical stresses or by chemical modifications of the surface, for instance in contact to SBF.

Furthermore, no thermal poling studies on *bioabsorbable* glasses, like soda-lime phosphates, were carried so far. This type of glass is degraded and absorbed in the human body when implanted in bone fractures (fracture fixation applications) [16,17]. Therefore, in the present study, we have chosen a $46.4\ SiO_2 - 25.2\ Na_2O - 25.2\ CaO - 3.2\ P_2O_5$ (46S4) bioactive glass and a $25\ Na_2O - 30\ CaO - 45\ P_2O_5$ bioabsorbable glass (Ca30 glass) for detailed studies of the electrical and electrochemical processes during thermal poling experiments.

## 2. Experimental

The bioactive glass $46.4\ SiO_2 - 25.2\ Na_2O - 25.2\ CaO - 3.2\ P_2O_5$ (46S4) and the bioabsorbable glass $25\ Na_2O - 30\ CaO - 45\ P_2O_5$ (Ca30) were prepared from $Na_2CO_3$, $CaCO_3$, $SiO_2$, and $(NH_4)_2HPO_4$ powders. Initially, the reagents $Na_2CO_3$ and $CaCO_3$ were kept in an oven and dried at 393 K in order to remove any traces of water and adsorbed gases. Then stoichiometric amounts of the starting materials were ground in an agate mortar for 45 min. The mixtures were placed in a Pt crucible and melted in an electric furnace; the 46S4 mixture at 1623 K for 2 hours and the Ca30 mixture at 1423 K for 1 hour. After complete homogenization, the melts were poured into preheated stainless steel molds with a cylindrical shape. The obtained bulk glass samples were then annealed 40 K below their respective glass transition temperatures (determined by differential scanning calorimetry) for 10 hours. The annealed glass discs were cut into 8-10 slices with a thickness of about 500 – 950 μm using a high-precision cutting machine (Struers Accutom–5) equipped with a diamond saw blade. Both faces of the glass slices were then polished by high-precision grinding using a lapping machine (Logitec PM5). Finally, we obtained sample thicknesses in the range from 400 to 900 μm. For the electrical experiments, Pt electrodes were sputtered on both faces of the samples.

The ac impedance analysis was carried out in a frequency range from 0.1 Hz to 1 MHz and at temperatures between 213 K and 673 K using a Novocontrol Alpha-AK impedance analyzer. The analyzer is equipped with a broadband high voltage amplifier providing a maximum voltage of 500 V (dc bias voltage + ac voltage amplitude). The sample temperature was controlled by the Novocontrol Quatro Cryosystem.

Thermally stimulated polarization/depolarization current measurements (TSPC/TSDC) were carried out using a Keithley 6517A electrometer controlled by the Novocontrol WinTSC 1.3 software. Polarization voltages up to 1 kV were applied in a temperature range from 213 K to 673 K, and polarization currents in a range from 1 fA to 20 mA could be detected.

Energy dispersive X-ray spectroscopy (EDX) was performed to obtain elemental concentrations throughout the cross-section of poled and depoled glasses, imaged with a scanning electron microscope (SEM).

## 3. Results
### 3.1. Isothermal conductivity and permittivity spectra of unpoled glass

The frequency-dependent conductivity $\sigma'(f)$ of 46S4 glass at different temperatures is shown in Fig. 1. At temperatures $\geq 283$ K, the conductivity isotherms are characterized by a low frequency-plateau and by a dispersive regime at higher frequencies. In the plateau regime, the conductivity is identical to the ionic dc conductivity $\sigma_{dc}$, caused by long-range transport of mobile $Na^+$ ions in the bulk of the glass. The dc conductivity of both glasses obeys an Arrhenius law with an activation energy of 0.84 eV for the 46S4 glass and 0.99 eV for the Ca30 glass, respectively.

The conductivity isotherm obtained at T = 573 K is characterized by a drop of the conductivity at frequencies below 1 Hz. This drop is caused by blocking of the $Na^+$ ions at the Pt electrode-glass interface resulting in the formation of space charge (electrode polarization). This space charge formation is also clearly visible in permittivity spectra $\varepsilon'(f)$ shown in Fig. 1. It leads to very large permittivities up to $10^8$ at low frequencies.

### 3.2. AC impedance spectra during thermal poling

Figure 2 shows ac impedance Nyquist plots for the Ca30 glass at different temperatures. The impedance spectra were taken at a bias dc voltage $V_{dc}$ = 490 V and a rms ac voltage $V_{ac}$ = 0.5 V. A heating rate of 0.75 K/min was used, so that the temperature change during the recording of a single spectrum (taking about 1 min) is very small.

At low temperatures, only one semicircle is found which is related to the bulk electrical properties of the glass sample. At temperatures above 500 K we observe the appearance of a second semicircle due to the formation of interfacial layers at the electrode-glass interfaces. The kinetics of this layer formation can be studied by recording spectra over an appropriate time interval at a constant temperature. As an example, spectra of the Ca30 glass at 513 K are shown Fig. 3. The conductivity spectrum of the unpoled glass is characterized by a low-frequency plateau. Poling with a dc voltage of 490V leads to a low-frequency dispersion reflecting the formation of the interfacial layers. After about 90 min, the low-frequency dispersion becomes almost independent of time, indicating that the interfacial layers do not grow further.

In the permittivity spectra, the interfacial layer formation leads to an additional dispersion with a low-frequency permittivity between $10^3$ and $10^4$. These values are much smaller than the permittivities due to electrode polarization. This indicates that the interfacial layers formed during poling are much thicker than the space charge layers caused by electrode polarization.

### 3.3. Thermally stimulated polarization/depolarization current measurements

The TSPC/TSDC spectra were obtained by the following approach. The sample was biased with a dc voltage $V_p$ and was heated with a constant rate. The currents induced by dipolar, conduction and electrochemical processes were recorded (TSPC measurements). Then, the sample was poled under dc bias $V_p$ at a temperature $T_p$ for a time interval $t_p$ = 45 min. Subsequently, the sample was quenched to 213 K under the applied dc voltage. Finally, the Pt electrodes were short-circuited, the sample was heated up with a constant rate, and the induced depolarization currents were recorded (TSDC measurement).

In Fig. 4 we compare the dc conductivity $\sigma_{dc}$ obtained from low-ac-voltage impedance measurements and the conductivity obtained from TSPC measurements (i.e. polarization current densities were converted into conductivities) for both glasses. At low temperatures, the low-voltage dc conductivity and the TSPC conductivity are virtually identical, showing that the polarization currents in this temperature range are exclusively due to $Na^+$ ion transport in the bulk. At temperatures of 433 K for the 46S4 glass and 493 K for the Ca30 glass, TSPC peaks occur. This implies that with increasing temperature, the progressive formation of resistive interfacial layers leads to a drop of the sample conductivity.

The results of the TSDC measurements on the 46S4 glass (depoling) are shown in Fig. 5(a). We find three depolarization current peaks, in the following called P1, P2 and P3. The activation energies, $E_a$, of the underlying dynamic processes were estimated using the following expression [18]:

$$\frac{E_a}{kT_{peak}^2} = \frac{\nu_0}{\beta} \exp\left[-\frac{E_a}{kT_{peak}}\right] \qquad (1)$$

Here, $\nu_0$ and $\beta$ denote the attempt frequency of the dynamic process and the heating rate, respectively. $\nu_0$ was assumed to be equal to the pre-exponential factor of the onset frequency $\nu^*$ of bulk conductivity dispersion [19] (i.e., $\sigma'(\nu^*) = 2\sigma_{dc}$). The estimated $E_a$ values for the three peaks are listed in Tab. 1.

The activation energy of P1 is similar to that of the dc conductivity. Therefore, we assign this peak to the depolarization of the bulk glass due to $Na^+$ ion transport. When low poling voltages were used, the peaks P2 and P3 were clearly separated on the temperature scale. The activation energies of these peaks are considerably higher than that of P1. These peaks are most likely due the annihilation of the interfacial layers under the electrodes, see discussion in Sec. IV. With increasing poling voltage, the peak magnitudes increase and the two peaks are not clearly separated anymore.

TSDC curves of the Ca30 glass poled under different conditions are shown in Fig. 5(b). Two peaks P1 and P2 are detected, and an additional peak at temperatures above 600 K, which is caused by the glass transition (Tg = 653 K). As for the 46S4 glass, the activation energy estimated for P1 is similar to that of the dc conductivity, while the activation energy of P2 is much higher, see Tab. 1.

The total electrical charge density $Q_t$ stored in the glass after poling can be calculated from the depolarization current $J_d(T)$ using the following equation [20]:

$$Q_t = \frac{1}{\beta} \int_{T_{min}}^{T_{max}} J_d(T)\, dT \qquad (2)$$

The integration limits $T_{min}$ and $T_{max}$ were chosen such that the relevant depolarization processes occur in this temperature window. $Q_t$ values obtained at different poling voltages and temperatures are shown in Tab. 1. As expected, the stored charge increases with increasing poling voltage and temperature. For comparison, we calculated a stored charge density $Q_{st}$ from the low-frequency capacitance of the sample in low-ac voltage impedance measurements. This implies that $Q_{st}$ is the charge density stored due to electrode polarization. We obtain values of $Q_{st}$ = 0.98 C/m$^2$ for the 46S4 glass and $Q_{st}$ = 0.35 C/m$^2$ for the Ca30 glass, respectively. These values are smaller than the $Q_t$ values obtained at high poling voltages and temperatures. Thus, the formation of the interfacial layers leads to more charge storage than simple electrode polarization.

### *3.4. SEM/EDX profiles*

The atomic concentration profiles of Na, Ca and O in a 46S4 glass close to the electrode-glass interfaces are shown in Fig. 6. The profiles were obtained by means of SEM/EDX. In the lower part of the figure, we plot the profiles after poling ($T_p$ = 473 K, $V_p$ = 500V, $t_p$ = 45 min) and in the upper part, we plot the profiles after depoling (heating rate 5 K/min up to 673 K and hold for 30 min). In the poled glass, we find an approximately 2 μm thick sodium depletion layer under the anode. On the other hand, the sample-cathode interfacial layer is characterized by a Na- and O-enriched layer followed by a layer with a slight Na depletion. It is important to note that there is no depletion or enrichment of Ca at the sample-electrode interfaces. This implies that Ca$^{2+}$ are immobile under the poling conditions. These experimental findings are summarized in a schematic illustration of the cross section of a glass sample with electrodes after poling, shown in Fig. 7.

After depoling the samples, the interfacial layers have disappeared completely, and the concentration of Na and O at the interfaces is virtually identical to the bulk of the glass, see upper part of Fig. 6.

## 4. Discussion

The SEM/EDX profiles indicate that the formation of the interfacial layers is exclusively caused by redox and transport processes of sodium. At the cathode-sample interface, Na$^+$ ions are reduced to metallic sodium. Metallic sodium is chemically unstable under ambient conditions and reacts readily with oxygen and water forming sodium oxide Na$_2$O and sodium hydroxide NaOH, respectively. These electrochemical and chemical reactions are responsible for the Na- and O-enriched layer. The layer with a distance from the cathode between about 1.5 μm and about 4.5 μm is characterized by a sodium depletion and is most likely a Nernst diffusion layer. At high poling voltages, the Na$^+$ ion concentration at the cathode-sample interface drops strongly, leading to the formation of the diffusion layer.

At the anode-sample interface, we find a Na depletion layer, but no oxygen depletion. An oxygen depletion layer would be expected if oxide ions in the glass were oxidized to gaseous O$_2$. Since the three-phase boundary between the sputtered Pt anode, the glass sample and the gas phase is very small, the charge transfer resistance for this oxidation should be very large. Therefore, the only process occurring at the anode-glass interface seems to be the electrically driven transport of Na$^+$ ions into the bulk of the glass, resulting in the Na depletion layer. However, it does not become clear from our present results why this depletion layer extends over approx. 2 μm. A simple electrochemical double layer should have an extension of the order of the Debye length of the Na$^+$ ions. Due to the high number densities of Na$^+$ ions, $N_V$, in the glasses ($N_V$ = 1.23x10$^{22}$ cm$^{-3}$ in 46S4 and $N_V$ = 8.14x10$^{21}$ cm$^{-3}$ in Ca30), the Debye length is of the order of atomic distances. From an application point of view, the large thickness of the Na depletion layer is of great importance, since it implies that the negative surface charge cannot be annihilated by removing a few atomic layers from the surface, for instances by means of mechanical stresses or chemical etching processes.

In the case of the 46S4 glass, the annihilation of interfacial layers during the depoling (TSDC experiment) results in two depolarization current peaks P2 and P3. The P2 peak is caused by the faster of the two processes with the lower activation energy. This suggests that the P2 peak is related to the annihilation of the Na depletion layer under the anode. This process involves exclusively transport of sodium ions and no charge transport across the electrode-glass interface. In contrast, the annihilation of the interfacial layer under the anode involves both charge transfer and sodium ion transport. Therefore, we think that the P3 peak is caused by this process. At present, additional experiments are underway in order to check this interpretation.

In the case of the Ca30 glass, one should also expect two depolarization current peaks P2 and P3. However, we find only one. A possible explanation is a similar activation energy of both annihilation processes due to a low charge transfer resistance between glass and cathode.

Finally, the question arises in what way the interfacial layers in the poled samples contribute to the additional semicircle in the low-ac voltage impedance spectra. The Na$_2$O/NaOH layer formed under the cathode is most likely a good ionic conductor and should, therefore, not lead to the additional semicircle. In contrast, the sodium depletion layer formed under the anode and the Nernst diffusion layer in the interfacial regime between bulk glass and cathode should be more resistive than the glass. Since the sodium depletion is more pronounced under the anode, we attribute the second semicircle mainly to this depletion layer. This hypothesis will be checked in the future by partial depoling of the poled samples and subsequent ac impedance analysis and SEM/EDX profiling.

## 5. Conclusions

After thermal poling, a sodium depletion layer is detected at the bioglass-anode interface, while a Na$_2$O/NaOH layer and a Nernst diffusion layer with sodium depletion are detected at the bioglass-cathode interface. The sodium depletion layer under the anode seems to be responsible for a second semicircle in the ac impedance spectra of the poled glasses. The annihilation of the interfacial layers during a TSDC experiments can lead to two depolarization current peaks. We tentatively attribute the peak with the lower/higher activation energy to the annihilation of the interface under the anode/cathode, but more experimental work is needed to check this.

We find that at the highest applied poling voltage of 750 V (corresponding to a poling field of 11.6 kV/cm), charge densities of about 3.5 C/m$^2$ are stored in the glass samples. Bioactivity tests are currently underway in order to find out if these charge densities are sufficient for an enhancement of the bioactivity.

**Acknowledgements**

One of us (C.R.M) thanks to the Alexander von Humboldt foundation for providing a research fellowship. Furthermore, we are grateful to Dr. S. Agarwal and Prof. A. Greiner, University of Marburg, for the possibility of differential scanning calorimetry measurements, and to Mr. M. Hellwig for the SEM/EDX profiling. Finally, we acknowledge critical reading of the manuscript by Dr. T.B. Adams.

**Figure captions:**

**Fig. 1.** Ac conductivity and ac permittivity spectra of 46S4 bioglass at different temperatures.

**Fig. 2.** Nyquist plots of ac impedance for Ca30 glass at different temperatures with dc bias voltage 490 V and rms ac voltage 0.5V.

**Fig. 3.** Time dependent (a) ac conductivity and (b) ac permittivity spectra of Ca30 glass at 513 K with bias dc voltage 490 V and rms ac voltage 0.5V.

**Fig. 4.** Conductivity of 46S4 glass and of Ca30 glass from thermally stimulated polarization measurements under high dc bias and from low-ac voltage impedance measurements.

**Fig. 5.** Thermally stimulated depolarization current density of (a) 46S4 glass and (b) Ca30 glass under different poling conditions.

**Fig. 6.** SEM/EDX profiles of Na, Ca and O concentrations in poled ($E_p$ = 11.6 kV/cm, $T_p$ = 473K, $t_p$ = 45min) and depoled 46S4 bioglass.

**Fig. 7.** Schematic illustration of the cross-section of a poled bioglass.

**Fig. 1.**

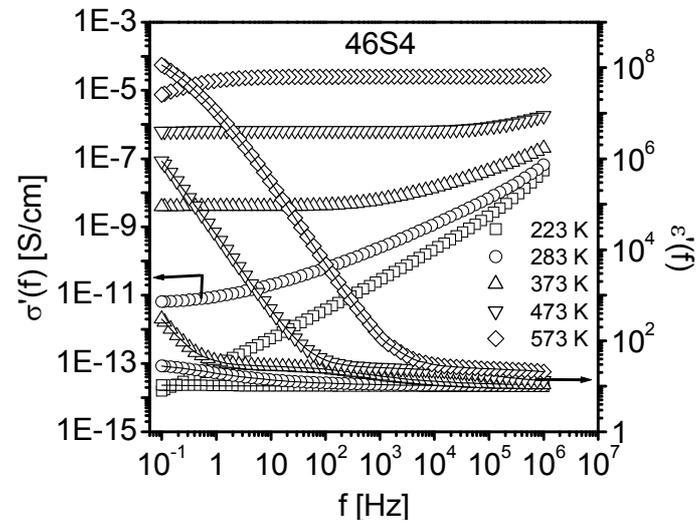

**Fig. 2.**

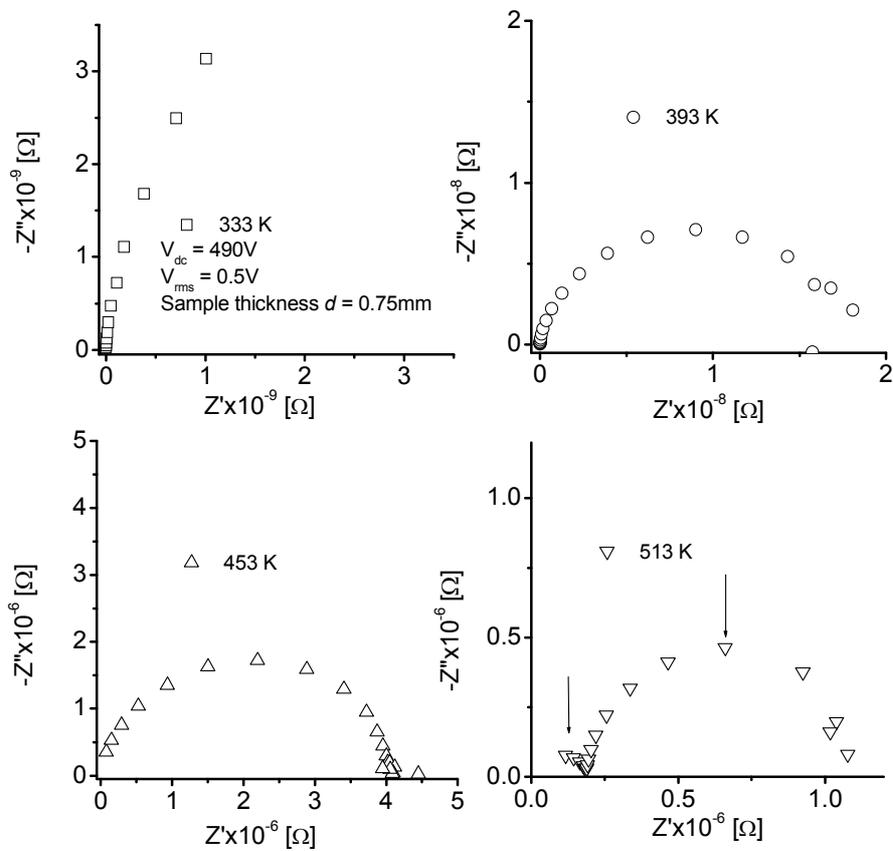

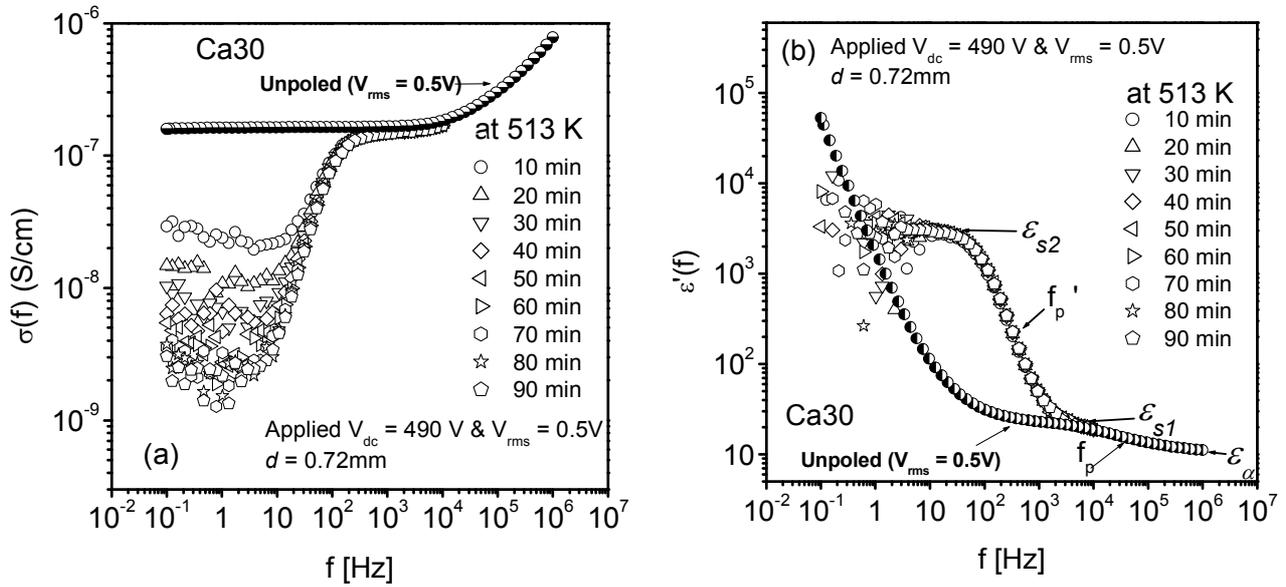

Fig. 3(a)&(b).

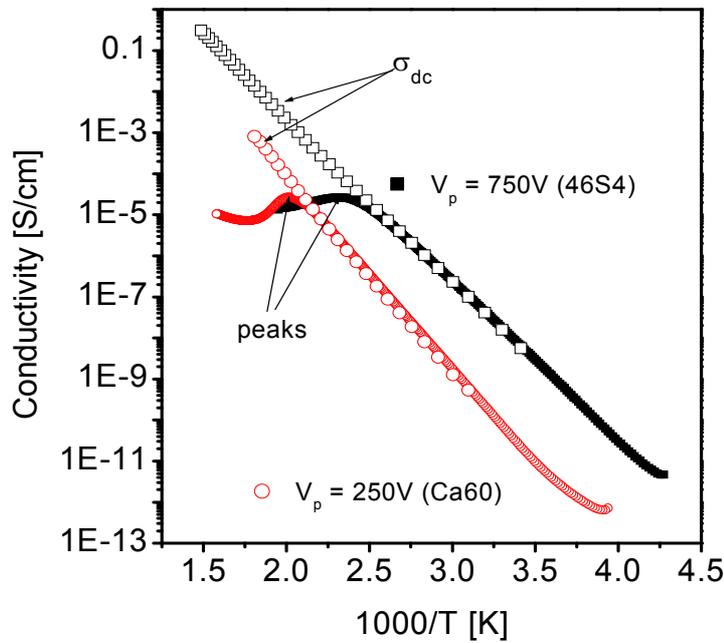

Fig. 4.

Fig. 5(a)&(b).

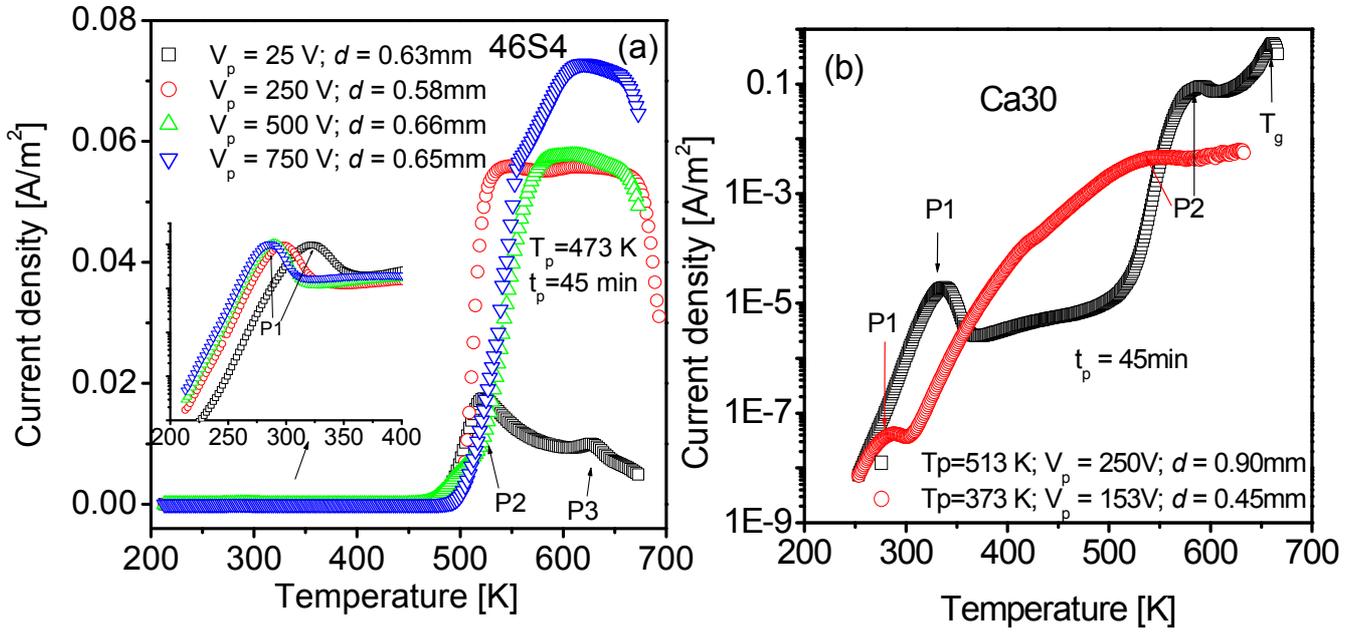

Fig. 6.

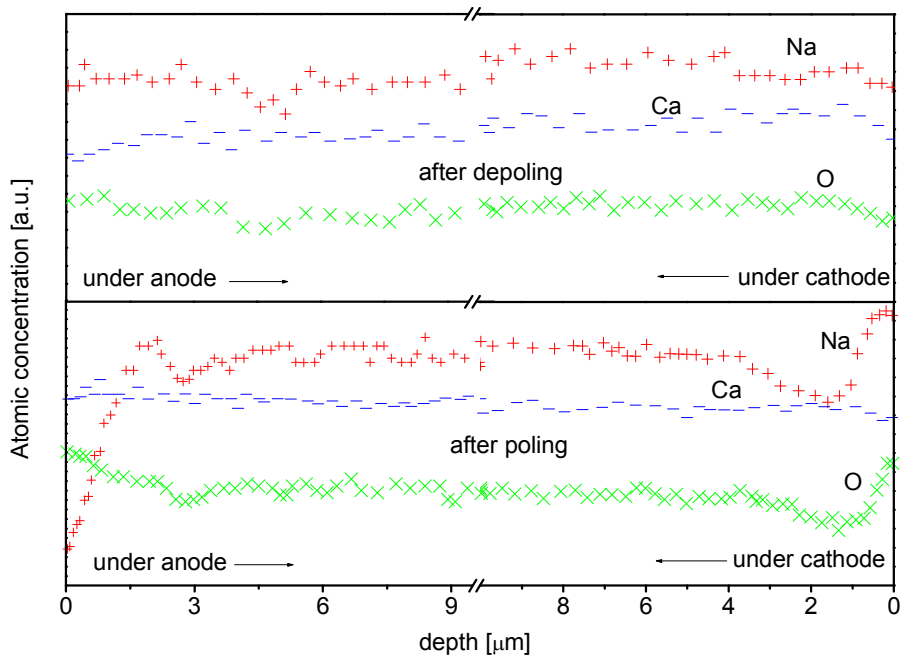

**Fig. 7.**

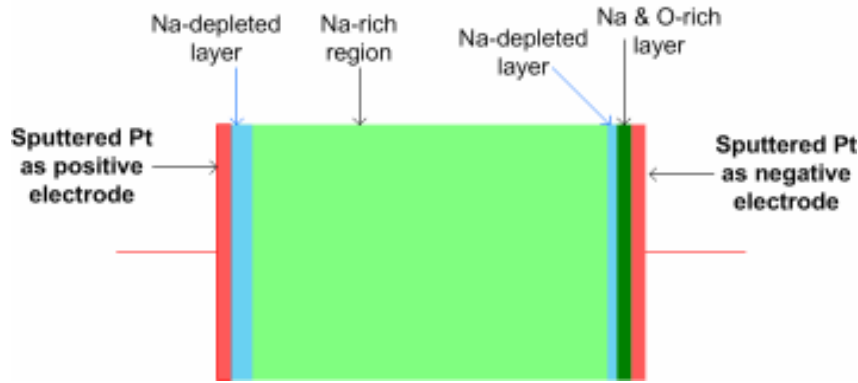

**Table 1.**

Effects of polarization field and temperature on total stored electrical charges ($Q_t$), activation energies ($E_a$) calculated from TSDC spectra, and the unpoled glasses the static surface charge ($Q_{st}$) and dc conductivity activation energy ($E_\sigma$) calculated from impedance data of 46S4 and Ca30 glass.

| 46S4 | | | | | Ca30 | | | |
|---|---|---|---|---|---|---|---|---|
| $E_p$ (kV/cm); $T_p$ (K) | $Q_t$ (C/m$^2$) | $E_a$ (eV) (from equ. (3)) | | | $E_p$ (kV/cm); $T_p$ (K) | $Q_t$ (C/m$^2$) | $E_a$ (eV) (from equ. (3)) | |
| | | P1 | P2 | P3 | | | P1 | P2 |
| 0.4; 473 | 0.49 | 0.86 | 1.60 | 1.92 | 3.4; 373 | 0.13 | 0.90 | 1.74 |
| 4.3; 473 | 2.12 | 0.88 | | | 2.7; 513 | 3.27 | 1.05 | 1.86 |
| 7.6; 473 | 2.90 | 0.89 | | | | | | |
| 11.6; 473 | 3.49 | 0.98 | | | | | | |
| Static surface charge, $Q_{st}$ | 0.98 C/m$^2$ | | | | Static surface charge, $Q_{st}$ | 0.35 C/m$^2$ | | |
| $E_\sigma$ (eV) (±0.02) | 0.84 | | | | $E_\sigma$ (eV) (±0.02) | 0.99 | | |